# Low-altitude aircraft will reshape noise exposure across global cities

Tianjing FENG[1], Jian KANG[1*]

[1] Institute for Environmental Design and Engineering, The Bartlett, University College London, London WC1H 0NN, United Kingdom

*Corresponding author(s). E-mail(s): j.kang@ucl.ac.uk

**Abstract**

Just as motorisation made road traffic a defining noise source of the twentieth-century city, low-altitude aircraft may reshape urban acoustic exposure in the twenty-first. Yet how this noise interacts with existing sound environments and three-dimensional urban form across global cities remains unclear. We modelled identical low-altitude aircraft operations across ten urban districts spanning all inhabited continents. Horizontal exposure varied markedly across and within cities, depending on road noise conditions and urban morphology, which also strongly influenced overall attenuation. Some semi-enclosed spaces with low noise levels before low-altitude aircraft operations experienced increases above 20 dB(A), creating pronounced local contrasts in urban noise exposure. Vertically, exposure varied with height and acoustic visibility to the flight route, creating vertical exposure inequalities between storeys, façades and buildings at comparable heights. These findings show that low-altitude aircraft can reshape noise exposure across global cities, requiring route assessment to distinguish newly exposed from already exposed areas and to account for three-dimensional exposure.

# 1 Introduction

Low-altitude aviation is developing rapidly worldwide. NASA frames Advanced Air Mobility as a future system for passenger transport, cargo delivery and public service operations[1], while EASA addresses urban air mobility through regulation and societal acceptance[2]. China has incorporated the low-altitude economy into national development policy and established dedicated institutions to support the sector[3,4]. Similar initiatives are advancing in the United Kingdom[5], Japan[6] and the United Arab Emirates[7], alongside developments in drone logistics, electric vertical take-off and landing (eVTOL) services, airspace trials and supporting infrastructure in other countries. Low-altitude aviation is becoming a rapidly growing and globally competitive urban industry[8,9].

Environmental noise is a major urban health burden, with long-term exposure to road traffic, railway and aircraft noise associated with annoyance, sleep disturbance, cardiovascular and metabolic effects, and impaired wellbeing[10]. Acoustic comfort in urban environments is also influenced by visual and broader multisensory conditions[11,12]. Existing urban noise policy was developed mainly for roads, railways and conventional airport aircraft. Low-altitude aircraft would add repeated aerial sound above streets, residential façades, courtyards and public spaces[13].

Emerging urban low-altitude operations include both cargo delivery drones and passenger eVTOLs. In this study, these two categories are collectively referred to as low-altitude aircraft. As urban noise sources, low-altitude aircraft differ from road traffic, railways and conventional aircraft in source height, spectrum, directivity and temporal structure. Multirotor drones commonly produce prominent tones and high frequency broadband components, and current evidence suggests greater annoyance than road-traffic or conventional aircraft noise at comparable sound-pressure levels[14-16]. The passenger eVTOL configuration examined here combines distributed electric propulsion with vectored thrust, producing changes in sound level, spectral content and directivity across flight modes[17]. Repeated pass-by along concentrated flight corridors and scheduled routes could turn occasional flyover sound into a regular component of the urban sound environment. Considering both low-altitude aircraft captures contrasting operating scales, mission types and acoustic signatures within emerging low-altitude aviation.

Established transport noise already exhibits substantial spatial exposure inequalities across urban environments[18].

For road traffic, street layout and three-dimensional urban form create substantial exposure differences between streets, sidewalks, inner courtyards, façade orientations and building heights[19,20]. Changes in building and street geometry can redistribute noise between exposed and shielded locations, producing marked horizontal and vertical contrasts even under the same traffic conditions[21-23]. Conventional aircraft noise is similarly shaped by the built environment, with buildings, terrain and the position of urban structures affecting its spatial attenuation and distribution[24-27]. Low-altitude aircraft may reshape urban noise exposure across the city. Upper façades, courtyards and other areas sheltered from ground transport noise may receive new or greater exposure, altering horizontal and vertical exposure patterns.

Previous research has mainly addressed other aspects of low-altitude aircraft and urban noise. Conventional aircraft studies focus on airports and established flight paths, while low-altitude aircraft research emphasises source characteristics, annoyance, trajectory planning and airspace management. Urban noise mapping also remains focused on road and railway noise, leaving the spatial exposure patterns created by low-altitude aircraft in dense cities less examined. Therefore, we ask two questions:

1. How do low-altitude aircraft operations reshape horizontal patterns of outdoor noise exposure across urban space?
2. How does low-altitude aircraft-induced noise reshape vertical patterns of façade exposure across building heights and façade positions?

To address this gap, we selected central 2 × 2 km districts in ten cities, including London, Tokyo, Hong Kong, Shenzhen, Beijing, Sydney, Santiago, New York, Nairobi and Dubai. Cities were selected primarily from the top ten urban centres in each geographical stratum according to combined population and gross domestic product (GDP) rankings. Final selection also considered low-altitude aviation development, urban morphology and geographical coverage. The simulations represented these two operational classes, with the DJI Matrice 600 representing urban delivery drones and the Joby representing passenger eVTOLs. Exposure was assessed on horizontal receiver planes and vertical façade grids to identify recurring and city specific three-dimensional patterns.

To avoid repetition, low-altitude aircraft are hereafter referred to as L-aircraft unless otherwise specified. Conventional aircraft refers to traditional aircraft operating from airports.

# 2 Results

The analysis covered ten 2 × 2 km urban districts: Beijing, Dubai, Hong Kong, London, Nairobi, New York, Santiago, Shenzhen, Sydney and Tokyo. Fig. 1 shows these districts differed strongly in building height, development intensity, building coverage and urban grain. 90th percentile (P90) building height ranged from 6.02 to 99.4 m, building footprint ratio from 0.25 to 0.51, floor area ratio (FAR) from 0.55 to 5.53, and building count from 602 to 16,461. These city districts covered four broad urban forms relevant to noise propagation: tower-dominated and discontinuous districts, continuous high- or mid-rise districts, fine-grained low-rise districts, and open low- to moderate-rise districts.

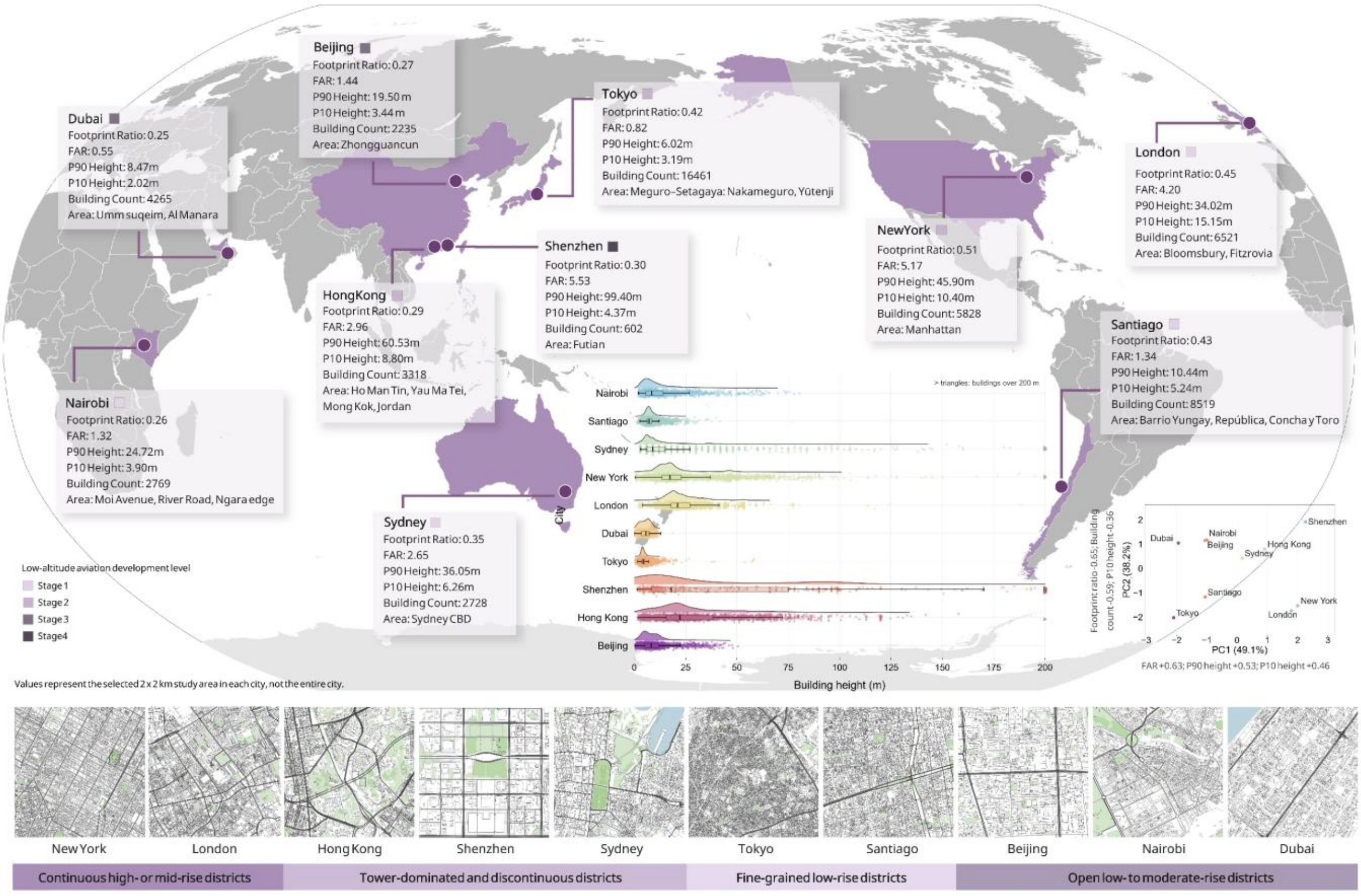

**Fig. 1 Global distribution and urban morphology of the selected 2 × 2 km study districts.** The coloured squares indicate low-altitude aviation development stage, and annotations summarise district morphology and location. Embedded plots show building-height distributions and an exploratory principal component analysis (PCA), in which PC1 (51.0%) represents vertical development intensity and PC2 (35.2%) contrasts open or discontinuous with fine-grained or continuous urban forms. The lower maps illustrate the four interpreted morphological configurations. The world map outline was adapted from *World Map 2*[28](FreeSVG, CC0); full data information is provided in Supplementary A Table 3.

## 2.1 Spatial distribution of L-aircraft-induced noise increases

Residents experience individual L-aircraft flyovers and fluctuations in road traffic. Day-evening-night sound level ($L_{\text{den}}$) summarises the cumulative exposure produced by these time varying events over the day, evening and night. $L_{\text{den}}$ at 4 m above local terrain was used as the metric for ground-level exposure.

Fig. 2 presents L-aircraft-only $L_{\text{den}}$ exposure from the modelled delivery drone and passenger eVTOL operations. Increasing source power and corridor flow expanded the affected area across both operational classes. Maximum L-aircraft-only $L_{\text{den}}$values occurred directly beneath the route, reaching approximately 65 dB(A) under the high-flow passenger eVTOL scenario, around 20 dB(A) higher than under the low-flow delivery drone scenario. Passenger eVTOL operations produced broader areas of high exposure across the district, particularly under Q150. Under low-flow delivery drone conditions, higher exposure remained more tightly concentrated around the flight corridor. The primary L-aircraft exposure footprint was concentrated beneath the flight corridor, with its extent increasing with source strength and corridor flow.

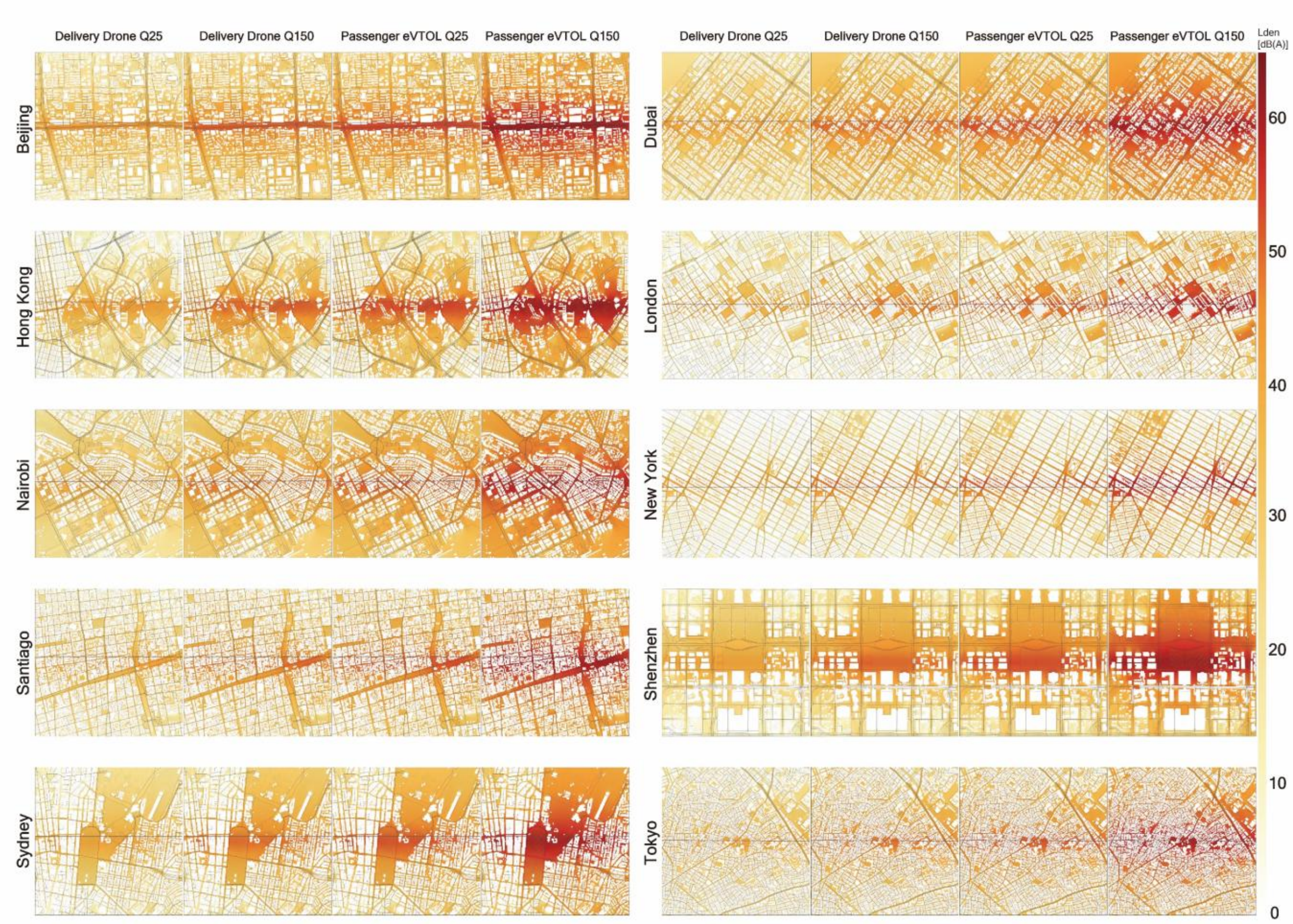

**Fig. 2 L-aircraft-only Lden at ground level.** L-aircraft-only represents noise attributable solely to L-aircraft operations. Base road-network and building-footprint data were based on OpenStreetMap extracts obtained via BBBike Extract (https://extract.bbbike.org/; © OpenStreetMap contributors, https://www.openstreetmap.org/copyright). Building heights, terrain data and city-specific building information were compiled from multiple public and published datasets; full data provenance, licences and coordinate reference systems are provided in Supplementary A Table 3.

Fig. 3 further shows that these inequalities become more pronounced when L-aircraft-induced $\Delta L_{\mathrm{den}}$ is evaluated against the existing road noise environment. In open low- to moderate-rise districts, including Beijing, Dubai and Nairobi, limited vertical obstruction produced relatively smooth and continuous footprints. The fine-grained low-rise districts in Santiago and Tokyo also showed broad ground-level exposure. Tower-dominated and discontinuous districts in Hong Kong, Shenzhen and Sydney showed localised exposed and shielded patches formed by towers and gaps. Roads, waterfronts, parks and other low-obstruction corridors provided clearer sound paths to the ground. Continuous high- or mid-rise districts in London and New York produced sharper contrasts between exposed corridors and sheltered spaces within urban blocks, reflecting the influence of urban morphology. Continuous building fronts and enclosed blocks restricted sound access to many internal locations, with openings, block edges and spaces with clearer overhead visibility showing larger local increases in noise level. Some inner-block spaces remained protected from road traffic while receiving L-aircraft sound from above, where L-aircraft-induced increases in noise level exceeded 20 dB(A) in several London and New York courtyards under the high-flow passenger eVTOL scenario. Increasing source power and corridor flow expanded the affected area across both operational classes, but patterns associated with urban morphology remained visible.

Across the ten cities, the magnitude of L-aircraft-induced change varied markedly under identical operations. Some cities showed little overall change, whereas others experienced much larger increases, creating a distinct cross-city exposure inequality. These local and cross-city contrasts show that existing road noise conditions can substantially reshape the magnitude and spatial distribution of aircraft-induced noise exposure.

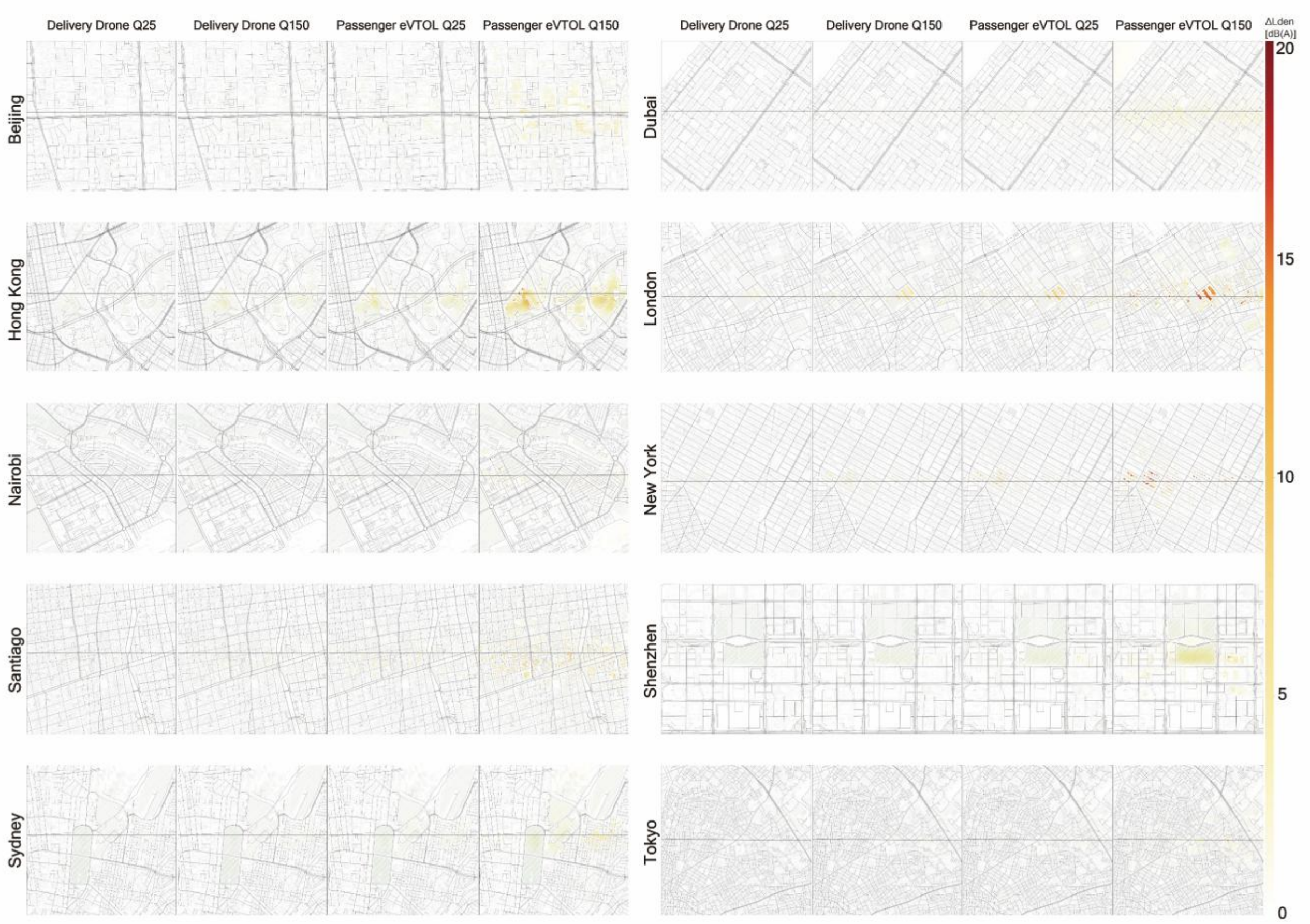

**Fig. 3 L-aircraft-induced ΔLden at ground level.** L-aircraft-induced $\boldsymbol{\Delta L_{\mathrm{den}}}$ represents the increase in dB(A) after L-aircraft noise is added to the road-traffic noise baseline. Base road-network and building-footprint data were based on OpenStreetMap extracts obtained via BBBike Extract (https://extract.bbbike.org/; © OpenStreetMap contributors, https://www.openstreetmap.org/copyright). Building heights, terrain data and city-specific building information were compiled from multiple public and published datasets; full data provenance, licences and coordinate reference systems are provided in Supplementary A Table 3.

Fig. 4 quantifies these spatial contrasts by examining the magnitude of L-aircraft-induced change across different existing road-noise levels. The same L-aircraft operations produced markedly different changes depending on where L-aircraft exposure overlapped with relatively quiet or already noisy urban spaces. These locations were frequently found in inner blocks, courtyards and other spaces shielded from road traffic but still exposed to sound from above, such as courtyards in London and New York. Under the eVTOL Q150 scenario, London had only a moderate L-aircraft-only footprint of 7.97%, but a relatively high newly exposed share of 2.11% and a $\Delta L_{\mathrm{den}}$ > 3 dB(A) share of 3.95%, while Santiago had a larger L-aircraft-only footprint of 12.52%, the largest newly exposed share of 3.78%, and the highest $\Delta L_{\mathrm{den}}$ > 3 dB(A) share of 5.5%. By contrast, locations along major roads, open corridors and building gaps often received greater L-aircraft exposure, yet the increase in total $L_{\mathrm{den}}$ was smaller because road noise already dominated the baseline. The combined Road + L-aircraft field remains dominated by the road network, with local levels approaching 90 dB(A) in several districts (Supplementary A Fig. 1). This explains why Tokyo, despite having an L-aircraft-only footprint of 12.47% under eVTOL Q150, produced only 0.36% newly exposed locations (see Supplementary A Table 5).

Under low-flow scenario (Q25), the lower source level produced little change across most road noise conditions, and under high-flow scenario (Q150) the larger increases remained concentrated mainly in relatively quiet locations. Passenger eVTOL scenarios extended these increases into locations with higher road noise, particularly under Q150. These results show that the added L-aircraft burden depended strongly on the existing road-noise baseline, producing large increases in quieter areas but little change where road noise already dominated, thereby reshaping the spatial pattern of noise exposure.

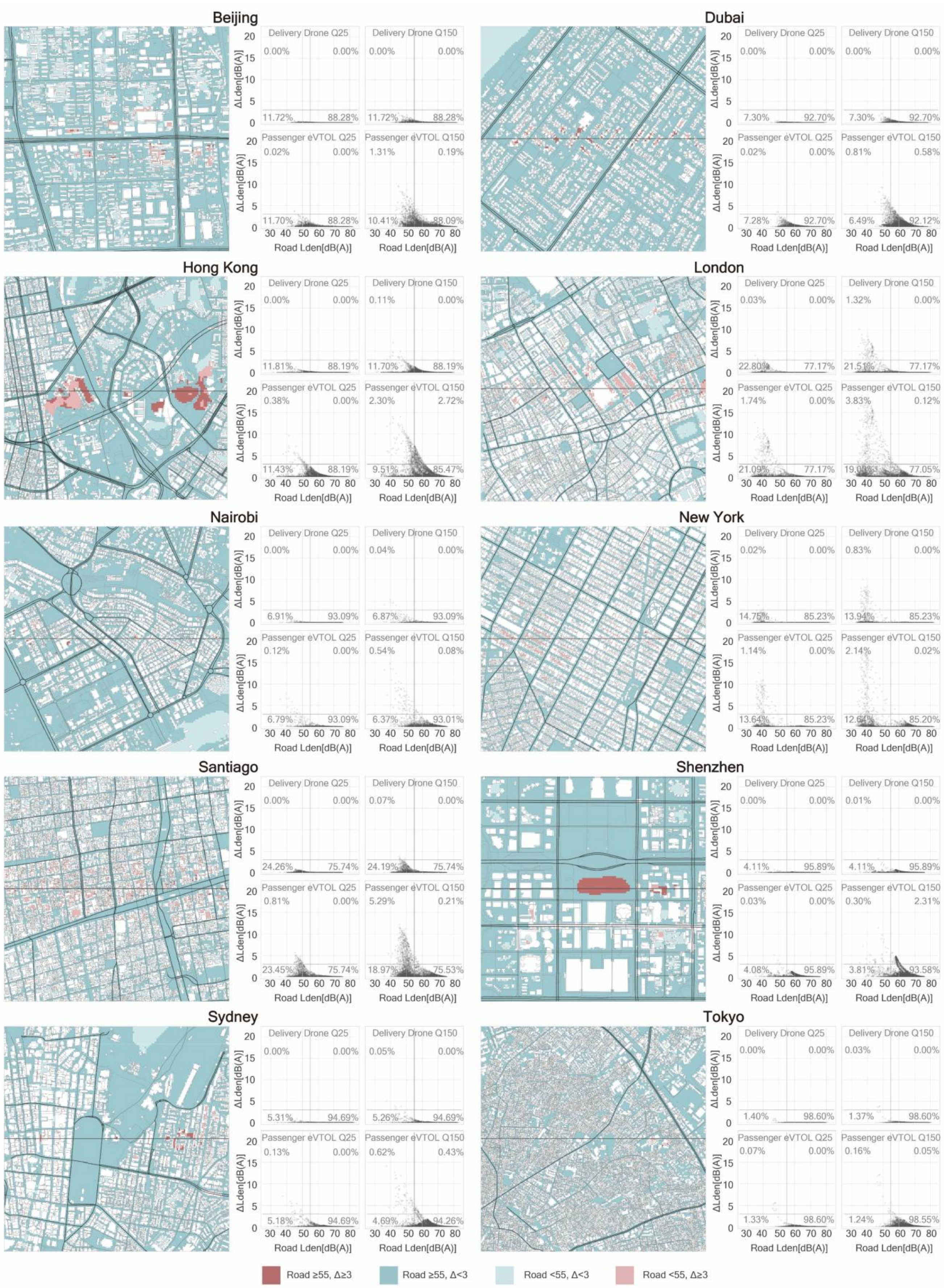

**Fig. 4 Relationship between road-only noise levels and L-aircraft-induced increases in noise level.** Each point represents a valid outdoor grid cell within one of the ten urban districts, where receiver located at ground level. The horizontal coordinate shows the Road-only day–evening–night sound level ($\boldsymbol{L}_{\mathbf{den}}$), and the vertical coordinate shows the L-aircraft-induced change, calculated for the same grid-cell location as $\boldsymbol{\Delta L}_{\mathbf{den}} = \boldsymbol{L}_{\mathbf{den,\ Road\ +\ Aircraft}} - \boldsymbol{L}_{\mathbf{den,\ Road-only}}$. The vertical reference line marks

a Road-only $\boldsymbol{L_{\mathrm{den}}}$ of 55 dB(A), and the horizontal reference line marks a $\boldsymbol{\Delta L_{\mathrm{den}}}$ of 3 dB(A), used as a general perceptibility criterion for environmental-noise change. Percentages indicate the share of valid outdoor grid cells within each region defined by the two reference lines. The plotted values were based on paired simulation outputs using identical receiver-grid coordinates across the Road-only and Road + L-aircraft scenarios. Road-network, building-footprint, building-height and terrain data were compiled from the sources listed in Supplementary A Table 3.

**Table 1** shows how far L-aircraft induced changes persisted from the route across cities. Noticeable increases faded within approximately 75 m of the route in Tokyo but remained detectable for 315–360 m in some cities, such as New York and London. More open urban structures and quieter receptor areas allowed the L-aircraft contribution to remain apparent farther from the corridor. High road noise backgrounds reduced the apparent influence of L-aircraft noise. The large variation in decay distance reveals substantial cross-city differences in exposure extent, with identical operations affecting markedly different spatial areas across cities. L-aircraft-only exposure, noticeable increases, newly exposed area, and decay distance capture complementary dimensions of horizontal exposure, indicating that a uniform route buffer would provide uneven levels of protection across cities.

**Table 1 Summary of L-aircraft noise contributions to 4 m horizontal outdoor exposure relative to road-only conditions.** All percentages refer to valid outdoor grid cells. Road-only, L-aircraft-only and Road + L-aircraft report the shares with $L_{\mathrm{den}} \geq 55$ dB(A). L-aircraft-induced change is calculated as $\Delta L_{\mathrm{den}} = L_{\mathrm{den,\,Road\,+\,L-aircraft}} - L_{\mathrm{den,\,Road-only}}$, with $\Delta L_{\mathrm{den}} > 3$dB(A) indicating a perceptible increase. Newly exposed area denotes cells increasing from below to at least 55 dB(A). For $\Delta L_{\mathrm{den}} > 3$dB(A) decay distance, "negligible" indicates cases where the peak affected-cell proportion is below 1%, so no decay distance is reported.

| City | L-aircraft | Flow Q | Road-only Lden >=55 dB(A) (%) | L-aircraft-only Lden >=55 dB(A) (%) | Road + L-aircraft Lden >=55 dB(A) (%) | $\Delta L_{\mathrm{den}}$ >3 dB(A) (%) | Newly exposed >=55 dB(A) (%) | Delta >3 dB(A) decay distance (m) |
|---|---|---|---|---|---|---|---|---|
| Beijing | Drone | 25 | 88.28 | 0 | 88.33 | 0 | 0.05 | negligible |
| | Drone | 150 | 88.28 | 0 | 88.51 | 0 | 0.23 | negligible |
| | eVTOL | 25 | 88.28 | 0.08 | 88.74 | 0.02 | 0.46 | negligible |
| | eVTOL | 150 | 88.28 | 11.81 | 90.63 | 1.5 | 2.35 | 285 |
| Dubai | Drone | 25 | 92.7 | 0 | 92.73 | 0 | 0.03 | negligible |
| | Drone | 150 | 92.7 | 0 | 92.91 | 0 | 0.21 | negligible |
| | eVTOL | 25 | 92.7 | 0 | 93.13 | 0.02 | 0.43 | negligible |
| | eVTOL | 150 | 92.7 | 13.22 | 94.29 | 1.39 | 1.59 | 225 |
| Hong Kong | Drone | 25 | 88.19 | 0 | 88.29 | 0 | 0.1 | negligible |
| | Drone | 150 | 88.19 | 0 | 88.59 | 0.11 | 0.4 | 30 |
| | eVTOL | 25 | 88.19 | 0.01 | 89.12 | 0.38 | 0.93 | 150 |
| | eVTOL | 150 | 88.19 | 13.37 | 91.26 | 5.01 | 3.07 | 210 |
| London | Drone | 25 | 77.17 | 0 | 77.19 | 0.03 | 0.02 | 45 |
| | Drone | 150 | 77.17 | 0 | 77.32 | 1.32 | 0.15 | 135 |
| | eVTOL | 25 | 77.17 | 0 | 77.42 | 1.74 | 0.26 | 165 |
| | eVTOL | 150 | 77.17 | 7.97 | 79.28 | 3.95 | 2.11 | 315 |
| Nairobi | Drone | 25 | 93.09 | 0 | 93.1 | 0 | 0.01 | negligible |
| | Drone | 150 | 93.09 | 0 | 93.15 | 0.04 | 0.05 | negligible |
| | eVTOL | 25 | 93.09 | 0 | 93.21 | 0.12 | 0.12 | 120 |
| | eVTOL | 150 | 93.09 | 8.14 | 93.72 | 0.62 | 0.62 | 270 |
| New York | Drone | 25 | 85.23 | 0 | 85.23 | 0.02 | 0 | negligible |
| | Drone | 150 | 85.23 | 0 | 85.23 | 0.83 | 0 | 135 |
| | eVTOL | 25 | 85.23 | 0 | 85.24 | 1.14 | 0.01 | 135 |
| | eVTOL | 150 | 85.23 | 6.11 | 85.73 | 2.16 | 0.51 | 360 |
| Santiago | Drone | 25 | 75.74 | 0 | 75.75 | 0 | 0.01 | negligible |
| | Drone | 150 | 75.74 | 0 | 75.86 | 0.07 | 0.12 | 75 |
| | eVTOL | 25 | 75.74 | 0 | 75.98 | 0.81 | 0.24 | 150 |
| | eVTOL | 150 | 75.74 | 12.52 | 79.52 | 5.5 | 3.78 | 330 |
| Shenzhen | Drone | 25 | 95.89 | 0 | 95.89 | 0 | 0.01 | negligible |
| | Drone | 150 | 95.89 | 0 | 95.91 | 0.01 | 0.02 | negligible |
| | eVTOL | 25 | 95.89 | 0.02 | 95.93 | 0.03 | 0.05 | 30 |
| | eVTOL | 150 | 95.89 | 15.6 | 96.19 | 2.61 | 0.3 | 90 |
| Sydney | Drone | 25 | 94.69 | 0 | 94.71 | 0 | 0.03 | negligible |
| | Drone | 150 | 94.69 | 0 | 94.78 | 0.05 | 0.09 | 90 |
| | eVTOL | 25 | 94.69 | 0 | 94.94 | 0.13 | 0.25 | 180 |
| | eVTOL | 150 | 94.69 | 15.2 | 95.87 | 1.05 | 1.18 | 225 |
| Tokyo | Drone | 25 | 98.6 | 0 | 98.61 | 0 | 0.01 | negligible |
| | Drone | 150 | 98.6 | 0 | 98.66 | 0.03 | 0.06 | 15 |
| | eVTOL | 25 | 98.6 | 0 | 98.71 | 0.07 | 0.11 | 15 |
| | eVTOL | 150 | 98.6 | 12.47 | 98.95 | 0.21 | 0.36 | 75 |

The observed patterns could not be represented adequately by any single morphology indicator. Taller buildings may shield ground-level locations but shift exposure towards gaps, route-facing edges or upper parts of the urban fabric.

Greater building coverage may reduce broad exposure while increasing contrasts between neighbouring spaces. Urban morphology shaped horizontal exposure inequalities by determining where additional exposure accumulated, either in quiet spaces shielded from road traffic or in existing noisy corridors.

## 2.2 Vertical Façade Exposure

Vertical façade exposure reveals a different form of inequality from that observed in the horizontal grid results, shaped by building height, façade orientation and acoustic visibility to the flight route. Hong Kong, New York and Shenzhen had P90 building heights of 60.53, 45.90 and 99.40 m, respectively, placing a substantial proportion of their upper façades closer to the 100 m flight route. By comparison, Tokyo, Dubai and Santiago had P90 heights of only 6.02, 8.47 and 10.44 m, leaving the route well above most of the local skyline. Fig. 5 reveals four characteristic forms of façade exposure inequality, including altitude peak, façade contrast, local amplification and noise intrusion.

Altitude peak describes exposure increasing with height towards the flight altitude before declining again (see Fig. 5a). Lower floors were often screened by front row buildings, podiums or surrounding blocks, with L-aircraft-only levels increasing once façades gained a clearer acoustic path to the flight route. For taller buildings, elevated exposure could extend across several floors around the 100 m flight altitude before declining with increasing vertical separation, a pattern clearly seen on the selected New York and Shenzhen façades extending above 200 m. The selected New York façade was also only about 10 m from the flight route and reached a maximum L-aircraft-only level of nearly 70 dB(A). Pronounced topographic relief in hilly cities can place entire buildings closer to the flight corridor. Even relatively low buildings may experience high L-aircraft noise exposure. The selected Hong Kong façade is less than 100 m tall, but the elevated terrain raises its roof elevation to approximately the same height as the flight corridor. These vertical differences were limited under the lowest L-aircraft noise scenario but became more pronounced as source level or corridor flow increased. This pattern creates exposure inequalities between storeys, with intermediate or upper floors sometimes more exposed than both lower and higher floors.

Façade contrast describes differences in exposure between the front, side and rear façades of the same building at comparable heights (see Fig. 5b). Façades with a clearer acoustic path to the flight route showed higher L-aircraft-only levels, while adjacent façades could be partly shielded by the building itself. In Sydney and Beijing, adjacent façades differed by about 3.5–4.0 dB(A) in mean level despite their similar overall distance to the corridor. The contrast was stronger in Shenzhen, where the maximum level reached 76.6 dB(A) on one façade and 69.3 dB(A) on the adjacent façade. The corresponding P90–P10 values were 10.5 and 4.9 dB(A). These contrasts show that façade orientation and acoustic visibility can create substantial exposure inequalities even at similar heights.

Local amplification refers to concentrated high exposure over limited sections of a façade associated with nearby buildings and narrow gaps (see Fig. 5c). Façade sections with more open surroundings generally showed weaker and smoother exposure, while nearby buildings and narrow gaps produced sharper spatial variation by changing acoustic visibility over short distances. In Sydney, the selected façade had the lowest mean level of the three cases at 48.4 dB(A), but levels still ranged from 43.9 to 55.2 dB(A), with higher exposure concentrated near the adjacent built mass. In Shenzhen, the narrow space formed by two closely spaced façades coincided with a localised amplification, with levels reaching 63.1 dB(A). In Nairobi, the central part of the façade lay close to a neighbouring building, while both sides opened into more exposed spaces. The façade map shows higher levels concentrated in this central section, with lower exposure towards the more open edges. This produces exposure inequalities within individual façades, with neighbouring positions experiencing markedly different noise levels.

Noise intrusion occurs where locations sheltered from road traffic remain exposed to L-aircraft noise from above (see Fig. 5d). The magnitude of $\Delta L_{\mathrm{den}}$ depended not only on L-aircraft exposure but also on the road noise baseline. In London and New York, the largest $\Delta L_{\mathrm{den}}$ occurred within more enclosed parts of the blocks, where surrounding buildings reduced the road noise baseline more strongly than the L-aircraft exposure. Hong Kong showed a broader and stronger intrusion pattern, with a mean $\Delta L_{\mathrm{den}}$ of 15.5 dB(A) and a maximum of 17.7 dB(A) across the façade. In this study, façade $\Delta L_{\mathrm{den}}$ values were lower than those observed at ground level, where increases exceeded 20 dB(A) in some locations, as façade receivers were generally more exposed to existing environmental noise than the most sheltered ground level locations. These results show that areas protected from ground transport noise can experience large L-aircraft induced changes, reshaping spatial patterns of noise exposure across otherwise sheltered urban spaces.

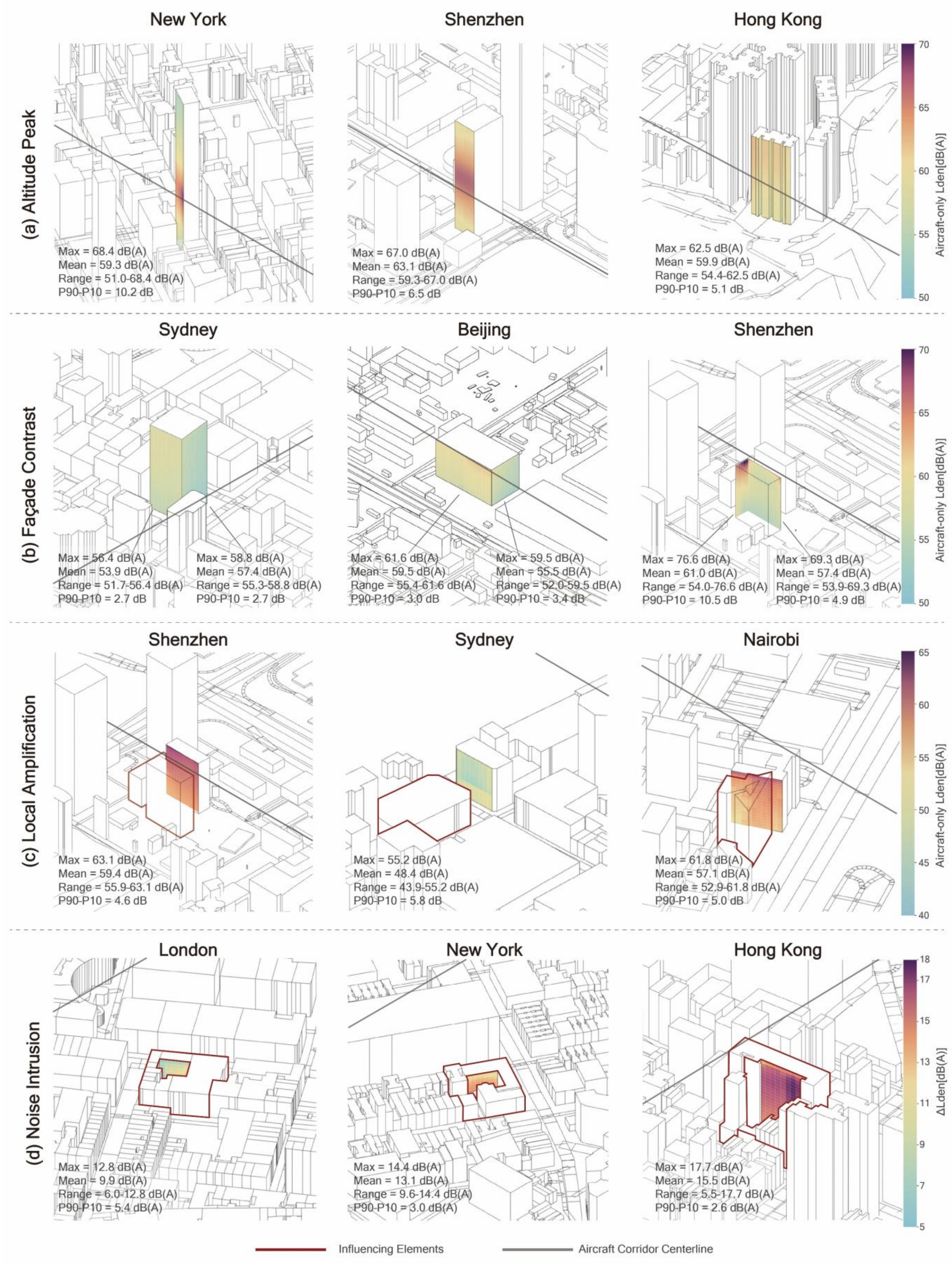

**Fig. 5 Representative façade patterns of L-aircraft noise exposure**. Examples show four characteristic spatial patterns observed across the urban districts: altitude peaks, façade contrast, local amplification and noise intrusion. Façade colours show L-aircraft-only $\boldsymbol{L_{\mathrm{den}}}$ for the first three rows and L-aircraft-induced $\boldsymbol{\Delta L_{\mathrm{den}}}$ for noise intrusion. Base road-network and building footprint data were based on OpenStreetMap extracts obtained via BBBike Extract (https://extract.bbbike.org/; © OpenStreetMap contributors, https://www.openstreetmap.org/copyright). Building heights, terrain data and city-specific building information were compiled from multiple public and published datasets; full data provenance, licences and coordinate reference systems are provided in Supplementary A Table 3.

To investigate how urban morphology relates to these exposure patterns across cities, Fig. 6 examines L-aircraft attenuation across frequency bands and its relationships with urban morphology metrics. Across the ten urban districts, octave band insertion loss varied substantially, while the frequency dependence of this attenuation was comparatively modest. In Fig. 6a, New York, London and Hong Kong showed consistently greater attenuation across most bands. Relative to the L-aircraft-only reference case containing no other modelled objects or sources, insertion loss (IL) in New York and London reached approximately 10–15 dB over much of the spectrum, compared with about 3–6 dB in most other districts. The apparent reduction in attenuation at 8 kHz resulted from strong atmospheric absorption under the adopted 10 °C condition, causing many receiver levels to approach the numerical floor, so this band was excluded from the spectral slope calculation.

Fig. 6b shows that P10 building height was the strongest correlate of attenuation, with Pearson coefficients of approximately −0.84 to −0.90 across 31.5–4000 Hz. FAR and footprint ratio ranked next, while P90 height was weaker and building count showed little association. These correlations are treated as exploratory because they are based on ten urban districts. In Fig. 6c, London, New York and Hong Kong had the three highest P10 values and also the greatest mean octave band attenuation, giving the strongest linear fit ($R^2$=0.79, RMSE = 1.39 dB). FAR ($R^2$=0.59, RMSE = 1.95 dB) and footprint ratio ($R^2$=0.46, RMSE = 2.23 dB) showed weaker fits, with cases such as Shenzhen combining high FAR with only moderate attenuation and Hong Kong showing strong attenuation despite a relatively low footprint ratio. A higher P10 is consistent with a more continuous urban canopy, suggesting stronger shielding than isolated tall buildings.

Fig. 6d separates attenuation magnitude from spectral reshaping. The spectral attenuation slope, $\beta$, measures the average change in attenuation for each octave increase in frequency between 31.5 and 4000 Hz. London, New York and Hong Kong generally showed more positive slopes, indicating preferential attenuation of higher frequencies, but most district medians remained below approximately 1 dB per octave. Urban morphology affected overall attenuation much more strongly than spectral shape.

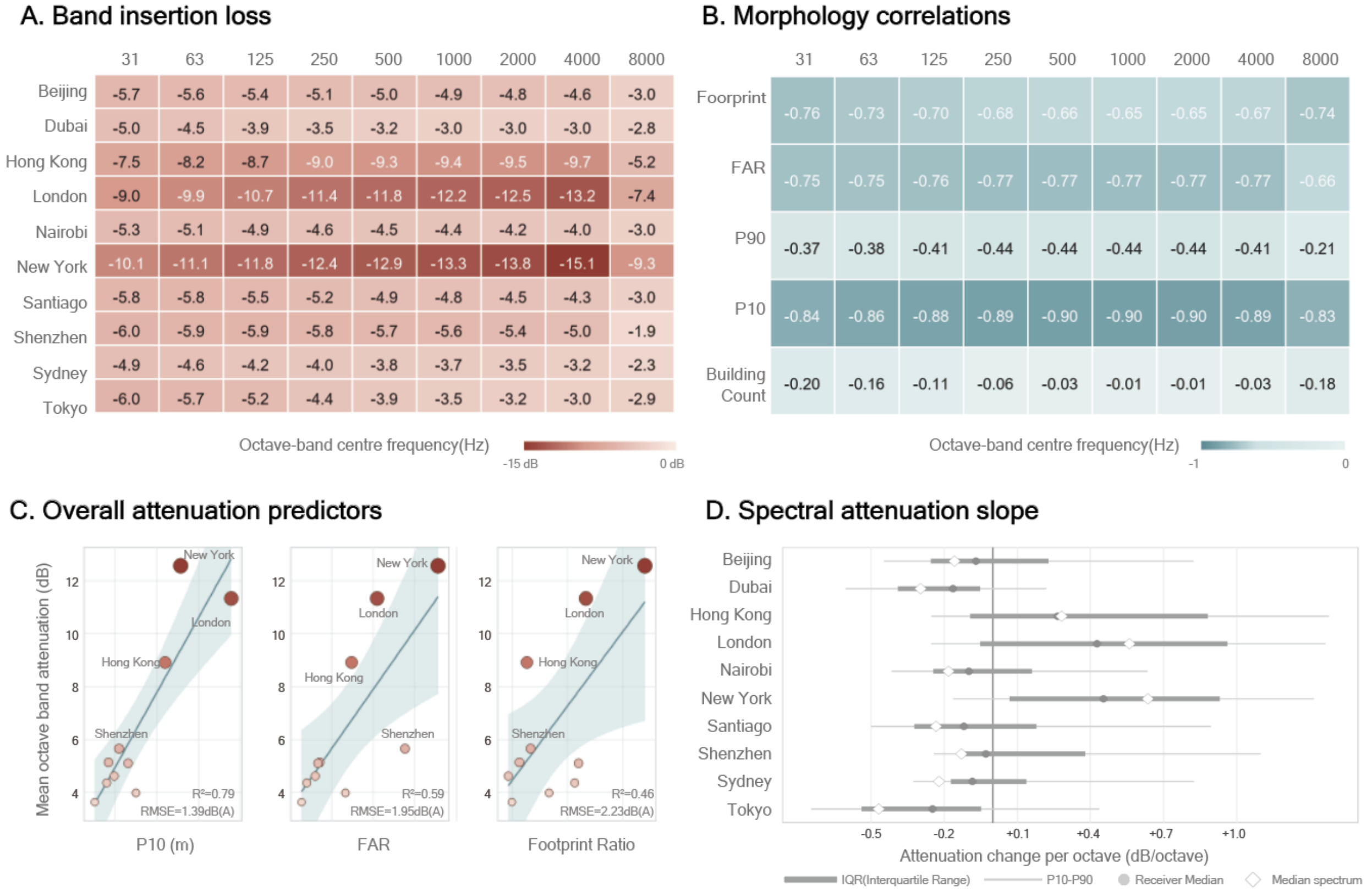

**Fig. 6 Urban morphology and L-aircraft noise attenuation.** a, Octave band insertion loss across cities; b, correlations between band insertion loss and urban morphology; c, associations between mean octave band attenuation and key morphology metrics, with 95% confidence intervals; d, attenuation change per octave across cities.

# 3 Discussion

Across ten urban districts spanning all inhabited continents, identical aircraft operations reshaped horizontal and vertical noise exposure patterns in markedly different ways. These differences reflected the combined effects of L-aircraft source strength, corridor flow, existing road noise conditions and urban form. Building shielding, acoustic visibility and baseline road noise conditions jointly reshaped L-aircraft exposure across urban space. Previous studies on how urban form affects L-aircraft noise have focused mainly on conventional aircraft near airports[24-26]. Other low-altitude aircraft research has concentrated on source characteristics[12,16], human response[13,15,25], operational parameters and route planning[11,23]. This study addresses this gap by providing a cross-city assessment of how L-aircraft noise interacts with existing road noise fields and three-dimensional urban form. Low-altitude aviation is moving from technology demonstration towards institutionalised development, regulation and early deployment across multiple world regions, making urban exposure assessment increasingly important. Comparison of real urban districts across multiple cities reveals how L-aircraft noise redistributes additional acoustic burdens across urban space, reshaping three-dimensional patterns of noise exposure.

**Noise management should distinguish newly exposed areas from already exposed areas experiencing low-altitude aircraft-induced increases.** Already noisy districts could have large L-aircraft-only footprints but little newly exposed area, whereas districts with lower pre-existing urban noise (e.g. noise from road traffic, construction and commercial activity) could show larger newly exposed shares and stronger L-aircraft-induced increases. Enclosed urban outdoor spaces were particularly vulnerable, with several London courtyards remaining shielded from road traffic but exposed to L-aircraft sound from above, producing increases exceeding 20 dB(A). This is consistent with evidence that L-aircraft noise produces smaller perceptual changes in noisy sound environments than in quieter ones[29,30]. However, aligning routes with busy transport corridors is insufficient because nearby road-shielded spaces may remain exposed from above. This contrast also extended across façade heights. Upper floors may retain direct acoustic paths to L-aircraft while lower floors benefit from urban screening, creating exposure patterns that cannot be inferred from ground level conditions alone. Conventional 4 m noise maps may underestimate both vertical exposure inequalities and

the extent of L-aircraft impacts within otherwise shielded urban spaces[31,32]. Low-altitude route assessment should combine horizontal noise maps, façade receivers and vertical sections while considering L-aircraft-only exposure, L-aircraft-induced change, newly exposed area and spatial extent of influence. Flight altitude, lateral offset, local morphology and pre-existing urban noise should inform route buffers and screening criteria[11].

**Urban design and route planning can use building configuration to reduce spatial exposure inequalities.** Across the ten districts, open low rise fabrics produced broader exposure, while tower gaps and continuous blocks created more localised exposed and shielded areas[33,34]. These patterns are consistent with traffic noise studies linking spatial attenuation to urban morphology[35] and L-aircraft studies linking attenuation to building frontal area, first row position, street topology, façade position and line of sight[24,26]. The selected courtyard façades differed by up to 12.2 dB(A), comparable with field measurements of up to 13.6 dB(A)[36] and drone modelling of up to 14 dB(A)[37]. Similar distances from a route can produce markedly different exposure due to urban form. Urban design and route planning can reduce spatial exposure inequalities by controlling how L-aircraft sound enters and propagates through urban spaces. Routes can avoid direct alignment with courtyard openings and major building gaps, while staggered openings can interrupt clear acoustic paths into sheltered spaces[38]. Absorptive treatment around courtyard openings and façades, or local acoustic screens, could further reduce sound entering or accumulating within enclosed areas[39]. Roof geometry provides another option[36], including slanted or retroreflective forms[40,41] that redirect incident sound away from sensitive spaces. These measures can reduce spatial exposure inequalities without simply transferring noise between neighbouring streets, blocks and buildings.

**Low-altitude aircraft noise assessment requires a three-dimensional perspective across receiver height and façade position.** Vertical exposure inequalities depended on receiver height relative to the 100 m flight route, surrounding rooflines and local geometry. Exposure often increased once façades cleared neighbouring obstructions, peaked near the flight-altitude band and declined above the route. This contrasts with conventional road exposure, which generally decreases above ground level[21]. Isolated buildings and buildings rising above surrounding roofs retained clearer acoustic paths than those embedded within continuous clusters, while route-facing façades were generally more exposed than side and rear façades. These results extend known effects of façade height, position, street topology and line of sight for conventional aircraft[26] to low-altitude routes and reveal a non-monotonic vertical exposure profile. Route planning should exploit urban morphology by using less exposed building sides and existing acoustic shielding, while avoiding direct passage along primary residential façades. Separation based on safety alone may not provide an acoustically appropriate setback, as previous studies have shown that distances between drones and façades may need to reach tens or even hundreds of metres[37,42]. In hilly cities, terrain should also be considered[43], as elevated ground can bring intermediate or upper floors closer to flight routes, as observed in Hong Kong.

**Spectral attenuation varied little across frequency, with most slopes below 1 dB per octave.** This may partly result from narrowband interference fluctuations being averaged within frequency bands, a pattern also observed in urban canyon modelling where one-third-octave attenuation showed little frequency dependence[44]. High frequency L-aircraft components, with their shorter wavelengths, may be more strongly attenuated by variations in urban morphology because diffraction becomes less effective as obstacle dimensions increase relative to wavelength. The ray tracing approach may also contribute to weak frequency dependence, as it is better suited to mid and high frequencies and represents low frequency diffraction less accurately.

Several limitations define the scope of these findings. The simulations represent level flyovers at a constant altitude of 100 m and use harmonised equivalent source spectra and assumed traffic flows. They do not capture take-off, landing, hovering, transition, L-aircraft directivity, meteorology or local operational demand. Each city is represented by one 2 × 2 km district, and road traffic inputs were standardised where detailed local data were unavailable. Future studies should test different route heights, lateral offsets, schedules, flight phases and vertiport locations across more cities and districts.

# 4 Methods

## 4.1 City Selection

The initial screening considered population and GDP, representing the scale of the potentially exposed urban population and the overall level of economic activity, respectively. To ensure comparability across countries, the analysis used the 2025 Global Human Settlement Layer (GHSL) World Urbanisation Prospects (WUP) Degree of Urbanisation (DEGURBA) urban-centre polygons[45]. DEGURBA is a harmonised methodology endorsed by the United Nations (UN) Statistical Commission that defines urban centres as contiguous 1 km² cells with population densities of at least 1,500 people per km² and a total population of at least 50,000[46]. This standardised definition reduces bias caused by differences in administrative area size and boundary-setting practices. Population data were obtained from the United Nations World Urbanization Prospects 2025 dataset [47], while 2024 gridded GDP values—the latest year available in the globally harmonised spatial dataset—were aggregated within the same urban boundaries [48,49].

Cities were assigned to geographical strata using the United Nations M49 classification [50], so that regional grouping followed an established international system. Population and GDP were ranked separately within each geographical stratum, and their mean rank was used as the Combined rank (see Supplementary B). Cities ranked within the top 10 in the combined ranking were retained, yielding 72 urban centres with high demographic and economic importance in their respective regional contexts.

The 72 candidates were assessed for evidence of urban logistics delivery and passenger eVTOL development in each city, the two low-altitude aviation applications examined in this study. Applications unrelated to the research question such as mapping, emergency response, policing, agriculture and medical-only logistics were excluded. Qualifying evidence was classified into four development stages. Stage 1 referred to a concrete city-level plan, formal agreement or approved preparatory action; Stage 2 to an actual flight demonstration or pilot; Stage 3 to limited but recurring operation on approved routes; and Stage 4 to routine commercial operation through an established network. Stage 0 was assigned where no qualifying evidence was identified for that city. Removing Stage 0 cities left 49 candidates. This threshold ensured that every selected city had at least a documented intention or practical action toward relevant low-altitude aviation. Early-stage cities including those with lower dominant GDP and population positions were deliberately retained as examples of future development where potential acoustic consequences could be examined before routine operations become established.

The final ten cities were selected from these 49 candidates through the comparison of urban morphology, geographical coverage and development stage. New York represents an extensive, continuous high-density and high-rise fabric. Three Chinese cities were retained because several Chinese candidates combined substantial urban scale and documented low-altitude aviation activity with markedly different urban forms. Shenzhen has a strongly vertical urban structure and established urban drone logistics, Hong Kong adds a compact high-rise environment despite falling below the combined population and GDP ranking threshold, and Beijing captures a predominantly mid- and low-rise superblock layout with emerging drone delivery and passenger eVTOL plans. Tokyo adds a polycentric East Asian megacity structure, with high-rise clusters embedded in a broad low- and mid-rise urban fabric. Dubai was selected for advanced air mobility infrastructure in a predominantly lower rise urban setting organised along major corridors. Nairobi was retained to include an emerging East African context. Sydney contributes an Australasian waterfront high-rise context with passenger eVTOL network preparation, and Santiago was included to capture an early-stage South American setting with a clear transition in surrounding mid- and low-rise districts. London was retained for the European stratum because detailed local knowledge supported site selection, input checking and interpretation. The ten cities span all inhabited continents and capture variation in aviation maturity, building height, building coverage, street structure, topography and the spatial concentration of high-rise development.

For each city, one 2 × 2 km district was delineated within the central urban area. These districts concentrate residents, employment, transport connections and multistorey buildings, creating plausible settings in which urban drone delivery and passenger eVTOL operations could interact with substantial exposed populations and complex built form. Each study district was selected where consistent building, road, terrain and traffic data were available. The characteristics of the selected districts are reported in Supplementary A Table 1.

Tower-dominated and discontinuous districts were most evident in Hong Kong and Shenzhen, with Sydney showing a milder form. Shenzhen combined the highest P90 height (99.40 m) with a moderate footprint ratio of 0.30, while Hong Kong had a P90 height of 60.53 m and footprint ratio of 0.29. All show strong vertical development dominated by tall buildings. Their low footprint ratios leave larger gaps between towers, creating a more open and discontinuous plan structure that reduces continuous shielding.

Continuous high- or mid-rise districts were represented by New York and London. New York combined FAR of 5.17 and footprint ratio of 0.51 with a P90 height of 45.90 m, while London had a lower P90 height of 34.02 m but a relatively high footprint ratio of 0.45. Both combine substantial vertical development with high plan coverage and relatively continuous building blocks. Compared with discontinuous tower districts, this more compact structure creates longer continuous building barriers and may provide stronger shielding from sound propagation.

Fine-grained low-rise districts were represented by Tokyo and Santiago. Tokyo had the lowest P90 building height (6.02 m) and the largest building count (16,461), while Santiago showed a similar combination of low building height and relatively high plan coverage. Both concentrate urban development in numerous closely spaced low-rise buildings, creating a dense and fine-grained plan structure. This form provides frequent but low building barriers, so sound propagation may be shaped more by repeated small scale obstruction than by tall building shielding.

Open low- to moderate-rise districts were represented by Beijing, Nairobi and Dubai. Their footprint ratios were consistently low, ranging from 0.25 to 0.27, while building heights and development intensity varied between the three cities. All combine relatively low plan coverage with wider spacing between buildings. The larger open areas and less continuous building structure may provide fewer barriers to sound propagation than the more compact urban forms.

The ten districts fall into four broad morphological types: tower-dominated and discontinuous districts, continuous high- or mid-rise districts, fine-grained low-rise districts, and open low- to moderate-rise districts. These combinations of height, coverage and urban grain provide the basis for interpreting the following noise-exposure patterns.

Urban model data were obtained from several open geospatial databases, including OpenStreetMap[51]. Detailed information on the data sources for each city is provided in Supplementary A, Table 3.

## 4.2 L-aircraft

Spectra from five L-aircrafts were reconstructed for comparative analysis. For the DJI Mavic 2 Pro and DJI S-900 [52], one-third-octave-band sound pressure levels were directly extracted from the reported spectra and used as the initial band level spectra. For the DJI Matrice 600 [53], T150 [54], and Joby eVTOL [17], the available spectra were reported as spectral density levels in dB/Hz.

L-aircraft noise can vary across operating conditions [13]. In this study, the Joby eVTOL and DJI Matrice 600 inputs used in the simulations were based on flyover spectral data. Ideally, the acoustic input data used for simulation should be obtained under the same flight speed and testing environment. However, research and practical deployment of industrial drones and eVTOL L-aircraft are still at an early stage [55], and publicly available spectral datasets remain limited.

For spectra reported in dB/Hz, values were extracted at the standard one-third-octave-band centre frequencies and converted into approximate band levels using a bandwidth correction. Specifically, the spectral density level at each centre frequency, fc, was assumed to represent the average spectral density within that band. The corresponding band level was then calculated as:

$$L_{p,\mathrm{band}}(f_c) = L_{p,\mathrm{dB/Hz}}(f_c) + 10\log_{10}(\Delta f)$$

where $\Delta f = f_{\mathrm{upper}} - f_{\mathrm{lower}}$ is the bandwidth of the one-third-octave band. The lower and upper band-edge frequencies were estimated as $f_c/2^{\frac{1}{6}}$and $f_c \times 2^{\frac{1}{6}}$, respectively [56]. This procedure converted the dB/Hz spectra into approximate one-third-octave-band levels, allowing all five L-aircraft to be represented using a common spectral format. The reconstructed spectra were used to compare the relative frequency distributions of the L-aircraft noise sources. In Fig. 7, each spectrum was normalised by its own maximum band level, with the highest band of each curve set to 0 dB. In these spectra, the heavier Joby eVTOL showed more low-frequency energy, while the lighter DJI Mavic 2 Pro showed stronger high frequency energy.

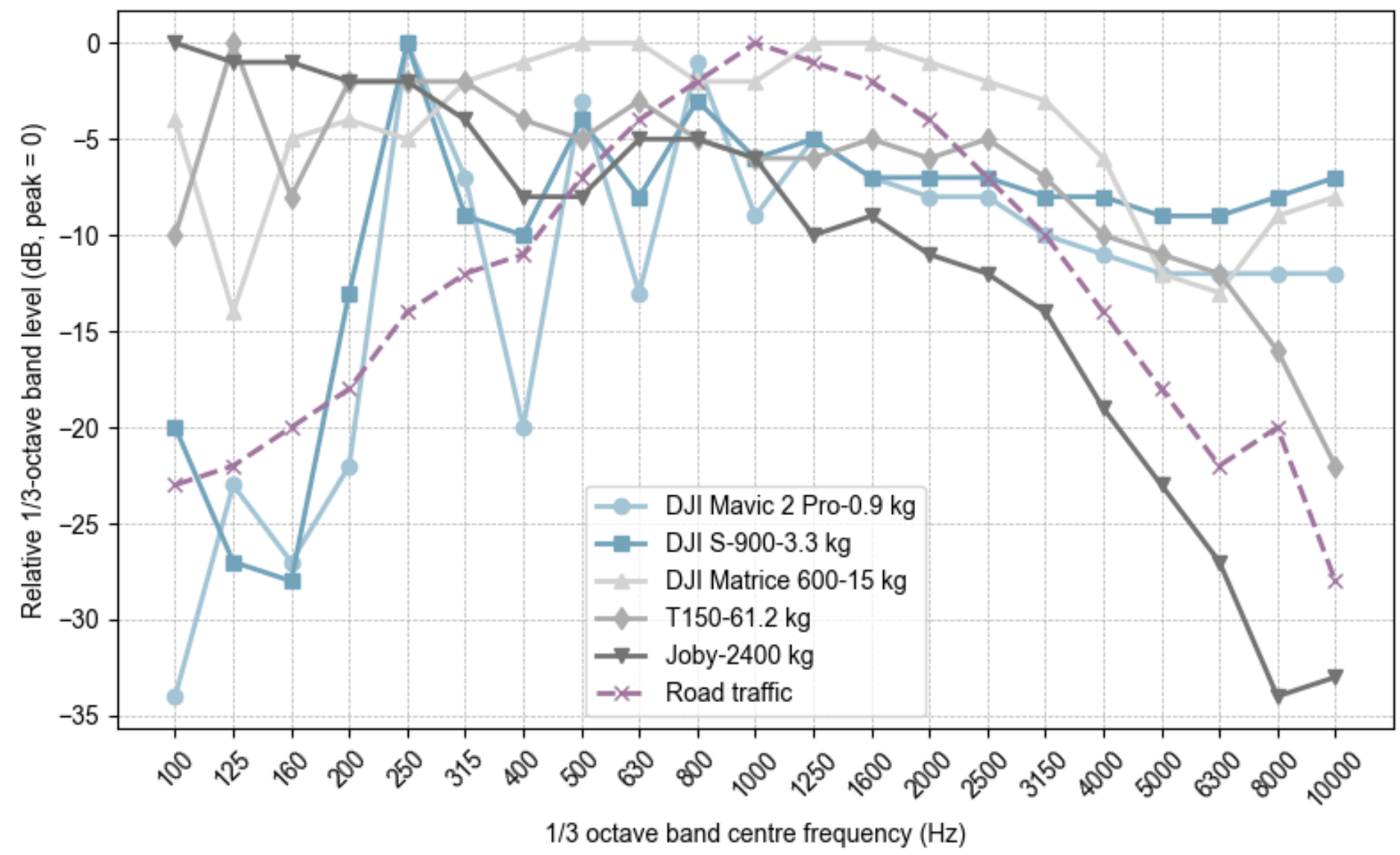

**Fig. 7 Normalised one-third-octave-band spectra of selected L-aircraft and road traffic.** Each spectrum was normalised to its maximum band level (0 dB), allowing comparison of spectral shape independent of absolute sound level. L-aircraft spectra were reconstructed from published data for the DJI Mavic 2 Pro and DJI S-900 [52], DJI Matrice 600 [53], T150 [54] and Joby eVTOL [17]. The road-traffic spectrum was based on published spectral data [57]. Spectral-density data were converted to one-third-octave-band levels as described in Methods. Because the original measurement conditions differed among sources, comparisons are limited to relative spectral shape.

The DJI Matrice 600 and Joby eVTOL were then selected as the two simulation sources. Because the original acoustic data were collected at different measurement or flyover distances, the measured sound pressure levels were normalised to a 1 m reference distance before defining the simulation inputs (see Supplementary A, Table 4). Under the assumptions of a point source, free field propagation and spherical spreading, the distance correction was calculated as:

$$L_w = L_p(r) + 10\log_{10}(4\pi r^2)$$

where $r$ is the measurement or flyover distance. The resulting 1 m reference levels were used as source strength proxies rather than direct final inputs. They were checked against the mass dependent multirotor emission trend, in which the 1 m emission level increases approximately logarithmically with take-off mass[52]. The final A-weighted Sound Power Level (LWA) inputs were set to 102 dB(A) for the DJI Matrice 600 and 112 dB(A) for the Joby eVTOL.

The scenario flow levels were defined by considering L-aircraft protected zones, minimum separation distances and published projections of future L-aircraft traffic flow. The protected zone of an L-aircraft is commonly simplified as a spherical, cubic, or cylindrical volume [58,59]. The protected zone was simplified as a cubic volume because the modelled L-aircraft were assumed to cruise without substantial roll or pitch and the spatial calculation was based on cubic grids.

Previous studies have defined L-aircraft separation using L-aircraft size multiples, temporal separation or fixed distance separation. Common fixed-distance values are generally concentrated between 10 and 30 m, while more conservative scenarios often adopt separation distances of 50 m or greater [60,61]. Time-based separation standards commonly range from 1 to 5 s [62], and cruise speeds of 10–15 m/s are frequently adopted in urban L-aircraft operation studies [58,63]. This study selected 15 m/s as the cruise speed and required the minimum L-aircraft separation distance to be no less than 10 m.

Research on future urban L-aircraft traffic levels remains limited, so approximate traffic scenarios were adopted. L-aircraft corridor flow intensities were selected from previous studies[64,65], scaled approximately to the 2 × 2 km modelling domain, and applied consistently across all L-aircraft types and urban districts to avoid introducing additional flow assumptions. The daytime scenarios were defined as low- and high-flow scenarios of 25 and 150 L-aircraft/h, respectively. The reference studies do not provide separate estimates for daytime, evening, and night time periods.

Evening flow was assumed to be 50% of daytime flow, giving 13 and 75 L-aircraft/h for the low- and high-flow scenarios. Night time operations were set to 0 L-aircraft/h, consistent with current commercial drone delivery assessments, which are generally restricted to daytime and evening hours[66].

## 4.3 Traffic

Motorways, expressways and other roads with grade separation and high speed limits were excluded from the modelling scope. For each urban district, the ordinary urban street network was divided into three road classes: L1 major urban arterials, L2 ordinary urban roads and L3 local two-lane streets.

The daily period was divided into three time periods: Day = 07:00–19:00, Evening = 19:00–23:00, and Night = 23:00–07:00, corresponding to 12 hours, 4 hours, and 8 hours respectively[31,32]. Where no traffic distribution by time period was available, the 24 hour AADT was split into D = 70%, E = 20% and N = 10%[67].

The traffic flows for each city were estimated by identifying available AADT or comparable traffic-count data for roads within or near the selected study urban district and converting these into hourly values. Detailed traffic profiles for day, evening and night periods were generally unavailable, so temporal disaggregation remained approximate. L1 roads show broadly comparable flows across cities, mostly around or above 2,000 vehicles $h^{-1}$ during the day, with larger inter-city differences observed for L2 and L3 roads. To improve cross-city comparability, road widths were standardised by road class, with L1, L2 and L3 roads assigned 6, 4 and 2 lanes in total in both directions, respectively, and a lane width of 3 m. Actual road widths vary within and between cities, but this simplified setting controls variation in road geometry and allows the simulations to focus more on differences in traffic flow, L-aircraft noise and urban morphology.

Speed limits were determined by local monitoring data, road classification information, or official traffic regulations where available. The speed limits for L1, L2 and L3 roads were set to 52.8, 44.7 and 34.4 km/h, respectively, based on the average values across the selected countries and regions. The specific numerical values and references are provided in Supplementary A Table 4.

## 4.4 Simulation Setting

Road-traffic noise was calculated in accordance with RLS-90[68]. The RLS-90 “No Foliage Attenuation” option was cleared to account for attenuation by green spaces and parks present in many cities. L-aircraft routes were represented as elevated line sources using octave-band spectra for each L-aircraft. Outdoor sound propagation was calculated using International Organization for Standardization (ISO) 9613[69,70], with air temperature and relative humidity fixed at 10 °C and 70% to ensure cross-city comparability. Buildings and terrain were represented in three dimensions, and route altitude was defined as a relative height above the local ground level. These calculations were implemented in CadnaA[71]. The default ground absorption factor was set to G=0, representing acoustically hard ground. Calculations used a maximum source search radius of 2,000 m and a maximum calculation error of 0.0 dB. All building façades were defined as acoustically reflective surfaces with a reflection loss of 1 dB[71].

The study-domain size was first justified using a simplified acoustic propagation calculation. Under the assumptions of a point source, free-field conditions, spherical spreading, no ground reflection, and no atmospheric absorption, the sound pressure level at distance r can be expressed as:

$$L_p(r) = L_w - 10\log_{10}(4\pi r^2)$$

For an L-aircraft with an A-weighted sound power level of $L_{\mathrm{WA}} = 112\mathrm{dB(A)}$, free-field spherical spreading predicts an A-weighted sound pressure level of approximately 45 dB(A) at a distance of 0.63 km. A 2 × 2 km domain centred on the flight corridor extends beyond the principal area expected to remain at or above this reference level under idealised free-field conditions. Atmospheric absorption, terrain effects and building shielding would generally reduce levels further. This domain size also captures a substantial range of surrounding urban form while avoiding the excessive computational cost associated with larger simulation areas.

Three source combination scenarios were modelled. Road-only included existing road traffic noise as the baseline,

L-aircraft-only included noise from the modelled L-aircraft operations, and Road + L-aircraft combined both source groups to represent the resulting urban noise environment. These scenarios isolated the acoustic contribution of L-aircraft and quantified their incremental effect relative to the road traffic baseline by retaining the same road network, building and terrain geometry, route alignment, receiver coordinates and calculation settings, with only L-aircraft type and corridor flow varied between scenarios.

Both the flight altitude and grid height were defined as relative heights. The cruise altitude was set to 100 m. Horizontal assessments were conducted at 4 m, following the European Union (EU) Environmental Noise Directive for strategic noise mapping[31,32], and at the maximum building height within each city plot. The 4 m results are referred to as ground-level maps. Horizontal noise maps were calculated using a 15 × 15 m grid, and vertical section maps used a 5 × 5 m grid to capture finer variations in noise propagation associated with urban morphology.

Façade receivers were generated along the selected building façades with a maximum horizontal spacing of 5 m. Each receiver was placed 0.1 m in front of the façade, with vertical receiver levels spaced every 3 m over the building height. Identical receiver coordinates were retained across Road-only, L-aircraft-only and Road + L-aircraft scenarios. To illustrate these propagation patterns more clearly, the Joby Q150 case was used for the façade visualisations, as it produced more pronounced vertical and lateral exposure contrasts. Façades were selected where nearby buildings, open spaces or enclosed street layouts produced clear differences in exposure across floors or along the façade. Maximum and mean values summarise façade receiver levels, while the range and P90–P10 difference characterise their spread.

In the absence of a validated health limit for L-aircraft noise, 55 dB(A) $L_{\mathrm{den}}$ was adopted as a screening threshold for sensitive and newly exposed areas, following World Health Organization (WHO) guidance and L-aircraft annoyance evidence[52,72]. A 3 dB(A) $\Delta L_{\mathrm{den}}$ criterion identified perceptible L-aircraft-induced changes[73]. The same $\Delta L_{\mathrm{den}} > 3$ dB(A) cells were then used to characterise spatial decay away from the route centreline. The decay distance was based on the full width at tenth maximum (FWTM) concept used in signal and imaging analysis. It was defined as the first distance band after the peak where the affected cell percentage fell below 10% of the scenario peak and stayed below this level in all subsequent bands[74].

# 5 Conclusion

Across ten urban districts spanning all inhabited continents, identical low-altitude aircraft operations produced markedly different noise exposure across and within global cities. Horizontal inequalities reflected the combined influence of existing road noise conditions and urban morphology, with some relatively quiet semi-enclosed spaces experiencing low-altitude aircraft-induced increases above 20 dB(A) despite modest district-wide changes. Urban morphology also strongly influenced overall low-altitude aircraft attenuation, although differences between frequency bands were comparatively limited. Vertically, exposure varied with height and acoustic visibility to the flight route, producing inequalities between storeys, façades and buildings at comparable heights. These findings demonstrate that low-altitude aircraft can reshape noise exposure across global cities in ways that cannot be captured by low-altitude aircraft-only maps or single-height assessments. Route assessment should distinguish newly exposed from already exposed areas and combine horizontal maps, façade receivers and vertical sections to account for three-dimensional exposure.

# Data availability

Building footprint and road network data were obtained from publicly available sources, including OpenStreetMap extracts downloaded via BBBike and city open data portals. The sources for each city are summarised in Supplementary A Table 3. The processed noise map grids and summary statistics supporting the findings of this study are available from the corresponding author upon reasonable request.

# Code availability

Custom Python scripts used for receiver coordinate matching, exposure metric calculation, urban morphology analysis and figure generation are available from the corresponding author upon reasonable request. CadnaA is proprietary software and is not redistributed. Model parameters and calculation settings are provided in the Methods and Supplementary A.

# Acknowledgements

We would like to thank colleagues in the University College London Acoustics & Soundscape Group for their constructive feedback during this study, with particular thanks to Vee Eawpanich. We also acknowledge that the CadnaA licence used in this study was provided by Huazhong University of Science and Technology, and thank Dr Hupeng Wu from Zhejiang Sci-Tech University and Dr Xinxin Li from Huazhong University of Science and Technology for their support.

# Authors and Affiliations

**Institute for Environmental Design and Engineering, The Bartlett, University College London, London WC1H 0NN, United Kingdom**

Tianjing Feng & Jian Kang

**Contributions**

T.F. and J.K. conceptualised the study and developed the methodology. T.F. conducted the software implementation, investigation, data curation, analysis and visualisation, and drafted the manuscript. T.F. and J.K. interpreted the results and reviewed and revised the manuscript. J.K. supervised and administered the overall project.

**Corresponding author**

Correspondence to Jian Kang.

**Symbols, abbreviations and definitions**

| Abbreviation | Full name | Unit |
|---|---|---|
| **NASA** | National Aeronautics and Space Administration | / |
| **EASA** | European Union Aviation Safety Agency | / |
| **UN** | United Nations | / |
| **EU** | European Union | / |
| **WHO** | World Health Organization | / |
| **ISO** | International Organization for Standardization | / |
| **GHSL** | Global Human Settlement Layer | / |
| **WUP** | World Urbanisation Prospects | / |
| **DEGURBA** | Degree of Urbanisation | / |
| **OSM** | OpenStreetMap | / |
| **UAV** | Unmanned Aerial Vehicle | / |
| **eVTOL** | electric Vertical Take-Off and Landing L-aircraft | / |
| **Q** | L-aircraft corridor flow | L-aircraft $h^{-1}$ |
| **Q25** | Low-flow scenario: 25 L-aircraft $h^{-1}$ during daytime, 13 L-aircraft $h^{-1}$ during evening, and no night time operations | L-aircraft $h^{-1}$ |
| **Q150** | High-flow scenario: 150 L-aircraft $h^{-1}$ during daytime, 75 L-aircraft $h^{-1}$ during evening, and no night time operations | L-aircraft $h^{-1}$ |
| **Lden** | Day-evening-night sound level | dB(A) |
| **ΔLden** | Change in day-evening-night sound level | dB(A) |
| **IL** | Insertion Loss | dB |
| **LWA** | A-weighted Sound Power Level | dB(A) |
| $L_w$ | Sound Power Level | dB |
| $L_p$ | Sound Pressure Level | dB |
| $R^2$ | Coefficient of Determination | Dimensionless |
| $\beta$ | Spectral attenuation slope | dB/octave |
| $G$ | Ground absorption factor | Dimensionless |
| **fc** | Centre frequency | Hz |
| **r** | Measurement or flyover distance | m |
| **FAR** | Floor Area Ratio | Dimensionless |
| **P90** | 90th percentile building height | m |
| **P10** | 10th percentile building height | m |
| **RMSE** | Root Mean Square Error | / |
| **FWTM** | Full Width at Tenth Maximum | / |
| **PCA** | Principal Component Analysis | / |
| **PC1** | First Principal Component | / |
| **PC2** | Second Principal Component | / |
| **AADT** | Annual Average Daily Traffic | vehicles/day |
| **D** | Day period traffic proportion | % |
| **E** | Evening period traffic proportion | % |
| **N** | Night period traffic proportion | % |
| **L1** | Road class 1 / major urban arterial roads | / |
| **L2** | Road class 2 / ordinary urban roads | / |
| **L3** | Road class 3 / local two-lane streets | / |
| **Road-only** | Road traffic only condition | / |
| **L-aircraft-only** | L-aircraft noise only condition | / |

| Road + L-aircraft | Combined road traffic and L-aircraft condition | / |
|---|---|---|
| L-aircraft-only Lden ≥ 55 dB(A) | Share of grid cells where L-aircraft-only noise exceeds 55 dB(A) | % |
| Road-only Lden ≥ 55 dB(A) | Share of grid cells where road-only noise exceeds 55 dB(A) | % |
| Road + L-aircraft Lden ≥ 55 dB(A) | Share of grid cells where combined noise exceeds 55 dB(A) | % |
| ΔLden > 3 dB(A) | Share of grid cells where L-aircraft noise increases total Lden by more than 3 dB(A) | % |
| Newly exposed area ≥ 55 dB(A) | Share of grid cells newly reaching or exceeding 55 dB(A) after L-aircraft noise is added | % |